\documentclass[10pt, conference, compsocconf]{IEEEtran}
\IEEEoverridecommandlockouts

\usepackage{graphicx}
\graphicspath{{figs/}}
\usepackage{latexsym}
\usepackage{amsmath}
\usepackage{amsfonts}
\usepackage{amssymb}
\usepackage[latin1]{inputenc}
\usepackage{color}

\usepackage{verbatim}

\usepackage{algorithm}
\usepackage{algorithmicx}
\usepackage{algpseudocode}

\algnewcommand\algorithmicparfor{\textbf{parfor}}
\algnewcommand\algorithmicpardo{\textbf{do}}
\algnewcommand\algorithmicendparfor{\textbf{end\ parfor}}
\algrenewtext{ParFor}[1]{\algorithmicparfor\ #1\ \algorithmicpardo}
\algrenewtext{EndParFor}{\algorithmicendparfor}

\usepackage{booktabs} 

\begin{document}
\title{Linear-Complexity Relaxed Word Mover's Distance with GPU Acceleration}

\author{\IEEEauthorblockN{Kubilay Atasu, Thomas Parnell, Celestine D\"unner, Manolis Sifalakis, Haralampos Pozidis, 
\\ Vasileios Vasileiadis, Michail Vlachos, Cesar Berrospi, Abdel Labbi}
\IEEEauthorblockA{IBM Research - Zurich\\
Zurich, Switzerland\\
\{kat,tpa,cdu,emm,hap,vva,mvl,ceb,abl\}@zurich.ibm.com}
}

\IEEEoverridecommandlockouts
\IEEEpubid{\makebox[\columnwidth]{978-1-4799-7492-4/15/\$31.00~
\copyright2017
IEEE \hfill} \hspace{\columnsep}\makebox[\columnwidth]{ }} 


\maketitle

\begin{abstract}
The amount of unstructured text-based data is growing every day. Querying, 
clustering, and classifying this big data requires similarity  computations across 
large sets of documents. Whereas low-complexity similarity metrics are available, 
attention has been shifting towards more complex methods that achieve a higher 
accuracy. In particular, the Word Mover's Distance (WMD) method proposed by 
Kusner et al. is a promising new approach, but its time complexity grows cubically 
with the number of unique words in the documents.
The Relaxed Word Mover's Distance (RWMD) method, 
again proposed by Kusner et al., reduces the time complexity from qubic to quadratic 
and results in a limited loss in accuracy compared with WMD. Our work contributes 
a low-complexity implementation of the RWMD that reduces the average time 
complexity to linear when operating on large sets of documents. Our linear-complexity 
RWMD implementation, henceforth referred to as LC-RWMD, 
maps well onto GPUs and can be efficiently distributed across a cluster of GPUs. 
Our experiments on real-life datasets demonstrate 1) a performance improvement of two 
orders of magnitude with respect to our GPU-based distributed implementation of the 
quadratic RWMD, and 2) a performance improvement of three to four orders of magnitude 
with respect to our distributed WMD implementation that uses GPU-based RWMD for pruning. 
\end{abstract}


\section{Introduction}

Modern data processing systems should be capable of ingesting, storing and searching
across a prodigious amount of textual information. Efficient text search
encompasses both a) high quality of the results and b) the speed of execution. 
Today, the quality of search results is guaranteed by a new class
of text representations based on neural networks, such as the popular
\textit{word2vec} representation~\cite{MikolovCorr2013}. The \textit{word2vec}
approach uses neural networks to map words to an appropriate vector
space, wherein words that are semantically synonymous will be close 
to each other. The Word Mover's Distance (WMD)~\cite{KusnerSKW15} 
proposed by Kusner et al. capitalizes on such vector representations 
to capture the semantic similarity between text documents. WMD is an 
adaptation of the Earth Mover's Distance (EMD)~\cite{RubnerTG98}, first 
used to measure the similarity between images. Like EMD, WMD constructs 
a histogram representation of the documents and estimates the cost of
transforming one histogram into another. The computational 
complexity of both approaches scales cubically with the size of 
the histograms, which makes their application on big data prohibitive.

To mitigate the high complexity of WMD, Kusner et al. proposed using a
faster, lower-bound approximation to WMD, called the Relaxed Word
Mover's Distance, or RWMD \cite{KusnerSKW15}. They showed that the
accuracy achieved by RWMD is very close to that of WMD. According to the
original paper, the time complexity of the RWMD method grows
quadratically in the size of the histograms, which is a significant
improvement with respect to the WMD method. However, the quadratic complexity
still limits the applicability of RWMD to relatively small datasets. 
In practice, computing pairwise similarities across millions of 
documents has extensive applications in querying, clustering, and 
classification of text data. This is precisely the focus of our work: to present a 
new algorithm for computing the Relaxed Word Mover's Distance that
 \textit{reduces the average time complexity from quadratic to linear}.
In practice, such a reduction in complexity renders the high-quality 
search results offered by WMD and RWMD applicable for massive datasets.
Additional \textbf{contributions} of this work include: 
\begin{itemize} 
\item A formal complexity and scalability analysis of the LC-RWMD method and other relaxations of WMD.
\item Massively parallel GPU implementations and scalable distributed implementations targeting clusters of GPUs.
\end{itemize}

\section{Word Mover's Distance} \label{sec:background}

The Word Mover's Distance assesses the semantic distance between two documents. 
If two documents discuss the same topic they will be assigned a low distance even if they have no words in common.
WMD consists of two components. The first is a vector representation of the words. 
The second is a distance measure to quantify the affinity between a pair histograms, wherein each histogram is a \textit{bag of words} representation of the respective document. 
The vector representation creates an embedding space, in which semantically synonymous words will be close to each other. 
This is achieved using \textit{word2vec}, which maps words in a vector space using a neural network. 
In fact, \textit{word2vec} uses the weight matrix of the hidden layer of linear neurons as the vector representation of the words from a given vocabulary. 

\begin{figure}[h!]
  \vspace{-0.1in}
  \centering%
  \includegraphics*[width=.6\linewidth]{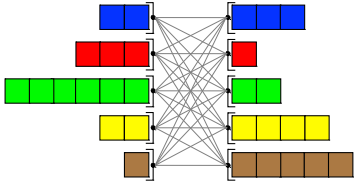}
  \vspace{-0.2in}
  \caption{Illustration of Earth Mover's Distance for two histograms}
  \label{fig:emd_illustration}
\end{figure}

The distance between two documents is calculated as the minimum cumulative distance of the words from the first document to those of the second document. 
This optimization problem is well studied in transportation theory and is called the Earth-Mover's Distance (EMD). 
Each document is represented as a histogram of words, and each word as a multi-dimensional vector, e.g., under \textit{word2vec} each word may be represented as a 300-dimensional vector. 
Fig.~\ref{fig:emd_illustration} depicts the computation of EMD between two histograms.

Given two L1-normalized histograms $x_1$ and $x_2$, where $\sum_px_{1}[p] = \sum_qx_{2}[q] = 1.0$, and a cost matrix $c$, EMD tries to discover a flow matrix $y$ that minimizes the cost of moving $x_1$ into $x_2$.
Formally, EMD is computed as follows:
\begin{align*}
& \text{EMD}(x_1,x_2) = \min \sum_{p,q}y[p,q]c[p,q] \quad \text{s.t.} \\
& y[p,q] \geq 0, \quad \sum_qy[p,q] = x_1[p], \quad \sum_py[p,q] = x_2[q],
\end{align*}
where $y[p,q]$ indicates how much of word $p$ in $x_1$ has to \textit{flow} to word $q$ in $x_2$, and $c[p,q]$ indicates the unit \textit{cost} of moving word $p$ in $x_1$ to word $q$ in $x_2$.
 
The combination of word embeddings and EMD forms the core of the WMD distance measure. 
In~\cite{KusnerSKW15}, WMD was shown to outperform seven state-of-art baselines in terms of the $k$-nearest-neighbor classification error across several text corpora. 
This is because WMD captures linguistic similarities of semantic and syntactic nature and learns how different writers may express the same viewpoint or topic, irrespective of whether they use same words.

\newcommand{\norm}[1]{\left\lVert#1\right\rVert}

\section{Complexity of WMD and Its Relaxations} \label{sec:complexity}

Assume that we are given two sets of histograms $X_1$ and $X_2$, and a vocabulary of size $V$. 
The sets $X_1$ and $X_2$ can be seen as sparse matrices, wherein each row is a sparse vector of dimension $V$. 
Each row $X[i]$ represents a histogram that is extracted from a text document and stores the weights (e.g., term frequencies) of the unique words in that document. 
The popular compressed sparse rows (csr) representation is a convenient way of storing the sparse matrices $X_1$ and $X_2$. 

\begin{figure}[t!]
  \centering%
  \vspace{-0.05in}
  \begin{tabular}{@{}c@{}}
  \includegraphics*[width=1.0\linewidth]{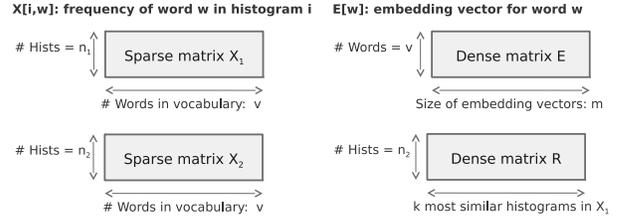}
  \end{tabular}
  \vspace{-0.2in}
  \caption{The sparse matrices $X_1$ and $X_2$ represent two document sets.} 
  \label{fig:notation}
\end{figure}

\begin{figure}[t!]
  \vspace{-0.05in}
  \centering%
  \begin{tabular}{@{}c@{}}
  \includegraphics*[width=1.0\linewidth]{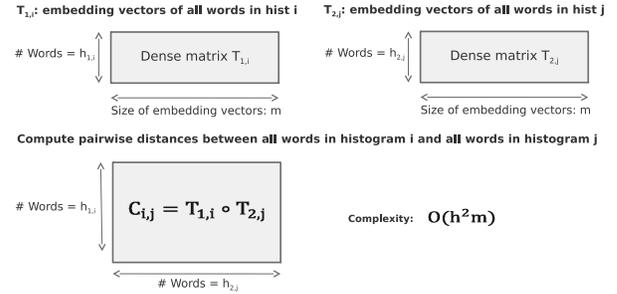}
  \end{tabular}
  \vspace{-0.25in}
  \caption{Given two histograms, the first step of WMD is to compute the Euclidean distances between all pairs of words across the two histograms.}
  \vspace{-0.10in}
  \label{fig:euclidean}
\end{figure}

Figure~\ref{fig:notation} shows the input and output data structures of the distance-computation algorithms. 
Suppose that the sparse matrices $X_1$ and $X_2$ have $n_1$ and $n_2$ rows, respectively. 
Assume that we are also given a dense matrix $E$, which stores embedding vectors for the words that belongs to a given vocabulary. 
For each word $w \in \{1..V\}$ in the vocabulary, $E[w]$ stores a vector of $m$ floating-point numbers given by some embedding process (\textit{word2vec} in the case of WMD). 
Given $X_1$, $X_2$, and $E$, our goal is to compute the distance between each pair $(X_1[i],X_2[j])$, $i \in \{1..n_1\}$, $j \in \{1..n_2\}$, and produce an $n_1 \times n_2$ matrix $D$ of distances.

Typically, a given histogram $X_2[j]$ is compared against all $X_1[i], i \in \{1..n_1\}$, which computes one row of $D$.
After that, the top-$k$ smallest distances and the positions of the respective histograms are computed in each row. 
The top-$k$ results for each row are then stored in an $n_2 \times k$ matrix $R$.

Given two histograms $X_1[i], i \in \{1..n_1\}$, and $X_2[j], j \in \{1..n_2\}$, the first step of WMD is to gather the vector representations of the words in $X_1[i]$ and $X_2[j]$ from matrix $E$. 
Assume that the number of nonzeros in $X_1[i]$ and $X_2[j]$ are $h_{1,i}$ and $h_{2,j}$, respectively. 
The dense matrix $T_{1,i}$ stores the vector representations of the words in $X_1[i]$ and has $h_{1,i}$ rows and $m$ columns.
Similarly, the dense matrix $T_{2,j}$ stores the vector representations of the words in $X_2[j]$ and has $h_{2,j}$ rows and $m$ columns.
Fig.~\ref{fig:euclidean} depicts this step.

The Euclidean distances between all pairs of word vectors in $T_{1,i}$ and $T_{2,j}$ form an $h_{1,i} \times h_{2,j}$ dense matrix denoted as $C_{i,j}=T_{1,i} \circ T_{2,j}$.
The $\circ$ operation is similar to a matrix multiplication between $T_{1,i}$ and the transpose of $T_{2,j}$, but instead of computing dot products between the word vectors, Euclidean distances are computed.
The complexity of the $\circ$ operation across two matrices, each with $O(h)$ rows and $O(m)$ columns, is the same as the complexity of a matrix multiplication operation and is given by $O(h^2m)$.

\begin{figure}[t!]
  \vspace{-0.25in}
  \centering%
  \begin{tabular}{@{}c@{}}
  \includegraphics*[width=1.0\linewidth]{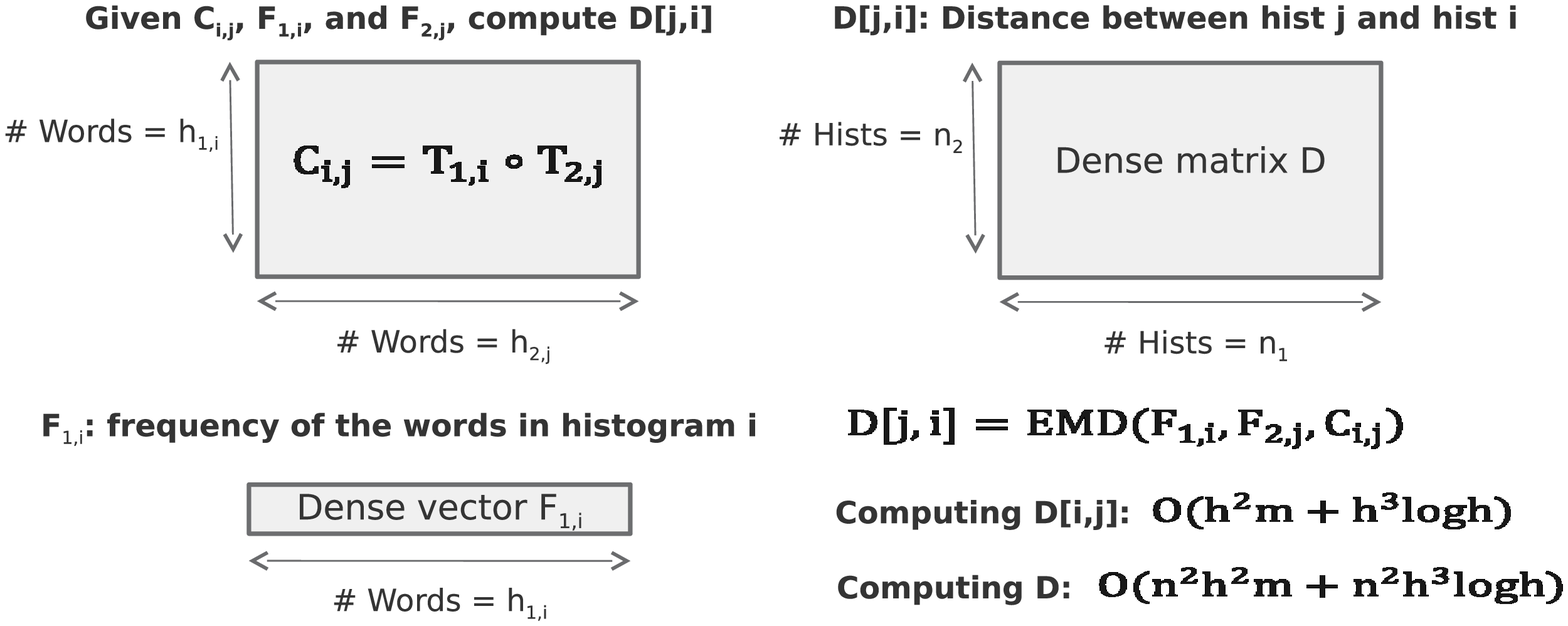}
  \end{tabular}
  \vspace{-0.3in}
  \caption{Application of EMD on pairwise distances to compute WMD.}
  \label{fig:wmd}
\end{figure}

\medskip
\noindent 
\textbf{Solving the WMD problem:}
Suppose that for each $X_1[i], i \in \{1..n_1\}$, a dense vector $F_{1,i}, i \in \{1..n_1\}$ is 
constructed in which only the nonzeros of $X_1[i]$ are stored. The size of $F_{1,i}$ is then $h_{1,i}$.
Similarly, suppose that for each $X_2[j], j \in \{1..N_2\}$, a dense representation $F_{2,j}, i \in \{1..N_2\}$ 
is constructed in which only the nonzeros of $X_2[j]$ are stored. The size of $F_{2,j}$ is then $h_{2,j}$.
EMD is computed based on $F_{1,i}$, $F_{2,j}$, and $C_{i,j}=T_{1,i} \circ T_{2,j}$ (see Fig.~\ref{fig:wmd}). 
The notation introduced is summarized in Table~\ref{tab:notation} and Table~\ref{tab:parameters}.

\begin{table}[t!]
\caption{Data structures used in WMD computation}
\vspace{-0.05in}
\label{tab:notation}
\begin{center}
\begin{tabular}{lcc}
  \hline
  Term          & Type  & Description \\
  \hline
  $E$           & Input & Embedding vectors for the vocabulary \\
  $X$           & Input & A set of histograms (csr matrix)\\ 
  $T_i$         & Auxiliary & Embedding vectors for histogram $i$ \\ 
  $F_i$         & Auxiliary & Term frequencies in histogram $i$ \\ 
  $D$           & Output & Distance matrix  \\ 
  $R$           & Output & Top-$k$ results  \\ 
  \hline
\vspace{-0.15in}
\end{tabular}
\end{center}
\end{table}

\begin{table}[t!]
\caption{Parameters used in WMD computation}
\vspace{-0.1in}
\label{tab:parameters}
\begin{center}
\begin{tabular}{lc}
  \hline
  $n$           & Number of histograms\\
  $v$           & Size of the vocabulary\\ 
  $m$           & Size of the embedding vectors \\ 
  $h$           & Size of a histogram \\ 
  $k$           & Used in top-$k$ calculation \\ 
  \hline
\vspace{-0.25in}
\end{tabular}
\end{center}
\end{table}

Assuming that the size of $F_{1,i}$ and $F_{2,j}$ is $O(h)$, the complexity of 
EMD computation is $O(h^3\log(h))$. Then, the overall complexity of WMD computation between two pairs of histograms is 
$O(h^2m+h^3\log(h))$. If $n_1$ and $n_2$ are both $O(n)$, the complexity of computing WMD across all pairs of histograms 
becomes $O(n^2h^2m+n^2h^3\log(h))$. So, the complexity of WMD grows quadratically in $n$ (number of documents) and cubically in $h$ (size of histograms).

\begin{figure}[t!]
  \vspace{-0.1in}
  \centering%
  \begin{tabular}{@{}c@{}}
  \includegraphics*[width=1.0\linewidth]{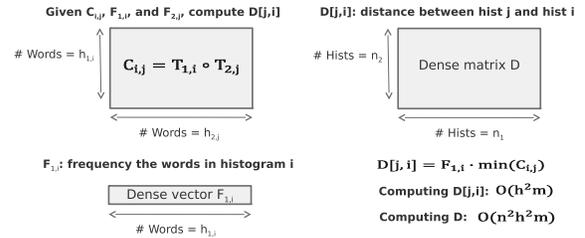}
  \end{tabular}
  \vspace{-0.45in}
  \caption{RWMD between histograms $X_1[i]$ and $X_2[j]$ is the dot-product of the vector $F_{1,i}$ and row-wise minimums of the matrix $C_{i,j}=T_{1,i} \circ T_{2,j}$.}
  \label{fig:rwmd}
\end{figure}

\medskip
\noindent
\textbf{Solving the RWMD problem:}
The RWMD provides a lower-bound approximation of the WMD.
Similarly to WMD, the first step of RWMD is the computation of $C_{i,j}=T_{1,i} \circ T_{2,j}$, which is a matrix of dimensions $h_{1,i} \times h_{2,j}$. 
The second step of RWMD is the computation of the minimum value of each row of $C_{i,j}$, which produces a floating-point vector of dimension $h_{1,i}$. 
The third and final step of RWMD is a dot-product operation between $F_{1,i}$ and the result of 
the second step, which produces a single floating-point value. Figure~\ref{fig:rwmd} illustrates the RWMD computation. 

Because the RWMD computation is not symmetric, it is advisable to perform it twice by
swapping $T_{1,i}$ with $T_{2,j}$ and $F_{1,i}$ with $F_{2,j}$, and by repeating the 
previous process. The symmetric lower bound is not necessarily equal to the first, 
and computation of the maximum of these two bounds provides an even tighter lower 
bound of WMD. In practice, it is not necessary to compute $T_{2,j} \circ T_{1,i}$ 
explicitly because this matrix is the transpose of $C_{i,j} = T_1[i] \circ T_2[j]$, which 
has already been computed.  It is sufficient to compute the minimum value in each column of 
$C_{i,j}$, and compute a dot product with $F_{2,j}$ to produce the symmetric lower bound.

Finally, the complexity of the overall RWMD computation is determined by the complexity of 
computing $C_{i,j}$, which is $O(h^2m)$. When computed across two sets of size $O(n)$ 
each, the overall complexity is $O(n^2h^2m)$, which is significantly lower than the 
complexity of WMD. However, the complexity of RWMD still grows quadratically with both 
$n$ and $h$, which can be impractical when $n$ and $h$ are large.

\medskip
\noindent
\textbf{Speeding-up WMD using RWMD:}
A pruning technique was presented in~\cite{KusnerSKW15} to speed up the WMD computation,
which uses RWMD as a lower bound of WMD. Given a histogram $X_2[j]$, first RWMD is 
computed between $X_2[j]$ and all histograms in $X_1$. Given a user-defined parameter 
$k$, the top-$k$ closest histograms in $X_1$ are identifed based on the RWMD distances. 
Next, WMD is computed between $X_2[j]$ and the top-$k$ closest histograms in $X_1$. 
The highest WMD value computed by this step provides a cut-off value $L$. WMD is 
computed between $X_2[j]$ and the remaining histograms in $X_1$ only if the pairwise 
RWMD value is lower than $L$. All other histograms will be pruned because they cannot 
be part of the top-$k$ results of WMD. A small $k$ leads to a small $L$, and
hence, a more effective pruning.

\medskip
\noindent
\textbf{Word Centroid Distance:}
Another approximation of WMD is Word Centroid Distance (WCD), wherein a single vector, 
called the centroid, is computed for each histogram $X[i]$ by multiplying $X[i]$ by $E$~\cite{KusnerSKW15}. 
This operation, essentially, computes a weighted average of all the embedding 
vectors associated with the words that are in $X[i]$. The WCD between two histograms $
X_1[i]$ and  $X_2[j]$ is given by the Euclidean distance between the respective centroids. 
The complexity of computing all centroids is then $O(nhm)$ and the complexity of 
computing all distances across two sets of size $O(n)$ each is $O(n^2m)$. When $n >> h$, 
the overall complexity becomes $O(n^2m)$. Although WCD has a low complexity, it is not a 
tight lower bound of WMD. Thus, it is not suitable when a high accuracy is required.

\section{Linear-Complexity RWMD} \label{sec:linearcomplexity}

A main disadvantage of the quadratic-complexity RWMD method is that it may compute the distances 
between the same pair of words $O(n^2)$ times. This overhead could be eliminated completely by precomputing the distances between all 
possible pairs of words in the vocabulary. However, such an approach would require $O(v^2)$ memory space, which typically is
prohibitive. Typical vocabulary sizes are on the order of a few million terms, which could, for instance, include company or person 
names in a business analytics setting. Storing distances across millions of terms requires tens of terabytes of storage space. 
In addition, accesses to this storage would be random, rendering software parallelization and hardware acceleration  impractical. 

Second, even if the distances across all words in the vocabulary are pre-computed and stored in a large and fast 
storage medium that enables parallel random accesses, if a given word $w$ appears in all histograms $X_1[i], i \in \{1..n_1\}$, 
the quadratic-complexity RWMD method would have to compute the word that is closest to $w$ in a given histogram $X_2[j]$ exactly $n_1$ times. 
Such a redundancy, again leads to a quadratic time complexity when this computation is repeated for all histograms $X_2[j], j \in \{1..n_2\}$.

We describe a new method for computing RWMD that addresses both of the aforementioned problems and reduces the time complexity to linear. 
Even though our approach may redundantly compute the distance between the same pair of words up to $n_2$ times, it entails a very low space 
complexity. In fact, our approach not only reduces the computational complexity with respect to the straightforward RWMD method, but also 
reduces the space complexity significantly. Because it requires a limited amount of working memory, it is very suitable for hardware 
acceleration. Lastly, our approach maps well onto the linear algebra primitives supported by modern GPU programming infrastructures. 

Reducing the time complexity of RWMD from quadratic to linear makes it possible to compute pairwise distances across 
1) large sets of documents (millions of documents) and 2) large histograms (millions of entries in histograms). 
We have implemented the new technique and showed that it is the only practical way of computing RWMD between all pairs 
across millions of documents, which leads to trillions of RWMD computations. Notably, we have measured a performance 
improvement of two orders of magnitude with respect to the quadratic-complexity RWMD method.

\begin{figure}[t!]
  \centering%
  \begin{tabular}{@{}c@{}}
  \includegraphics*[width=1.0\linewidth]{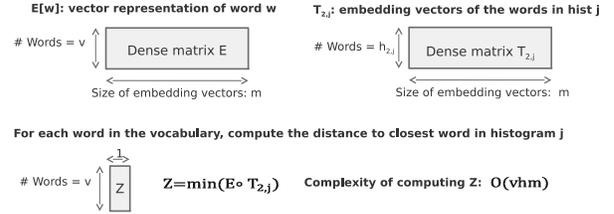}
  \end{tabular}
  \vspace{-0.6in}
  \caption{Linear-Complexity RWMD: First phase.}
   \label{fig:lcrwmd1}
\end{figure}

\begin{figure}[t!]
  \centering%
  \begin{tabular}{@{}c@{}}
  \includegraphics*[width=1.0\linewidth]{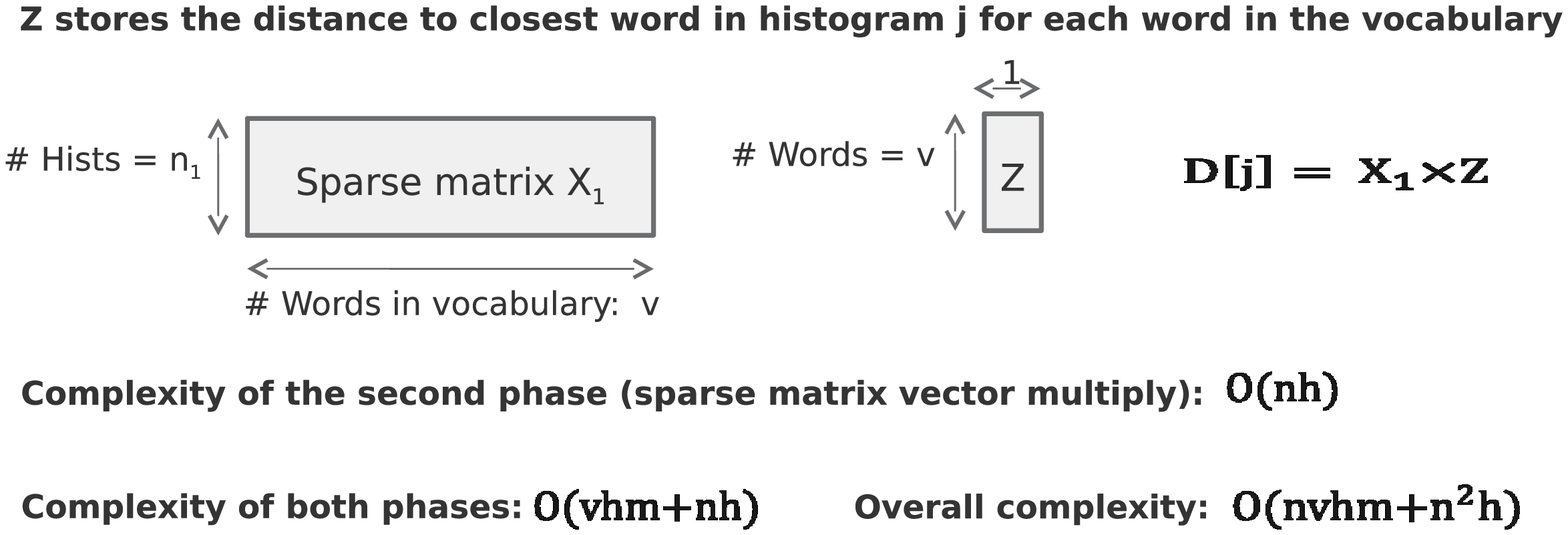}
  \end{tabular}
  \vspace{-0.7in}
  \caption{Linear-Complexity RWMD: Second phase.}
   \label{fig:lcrwmd2}
\end{figure}

The focal point of our work is a novel decomposition of the RWMD computation into two phases, with each phase having linear complexity 
in terms of the size of the histograms in $X_1$ and $X_2$. For a given histogram $X_2[j], j \in \{1..n_2\}$, the first phase computes 
the Euclidean distance to the closest entry in histogram $X_2[j]$ for each word in the vocabulary, which produces $Z$, a dense 
floating-point vector of dimension $v$ (see Fig.~\ref{fig:lcrwmd1} for illustration). The second phase performs a sparse-matrix 
dense-vector multiplication between $X_1$ and $Z$, which produces the RWMD between all $X_1[i], i \in \{1..n_1\}$ and $X_2[j]$ 
(see Fig.~\ref{fig:lcrwmd2} for illustration). 

The first phase is similar to the quadratic-complexity RWMD implementation given in Section~\ref{sec:complexity} in that 
it treats the vocabulary as a single histogram and computes pairwise Euclidean distances across all words in the vocabulary 
and the words in a given histogram $X_2[j]$ by computing $E \circ T_2[j]$. Next, row-wise minimums are computed to derive the 
minimum distances between the words in the vocabulary and the words in $X_2[j]$, and the results are stored in the vector 
$Z$ of size $v$. The complexity of this phase is determined by the $\circ$ operation and is given by $O(vhm)$. 

In the second phase, $X_1$ is multiplied by $Z$ to compute the RWMD. This phase is essentially a sparse-matrix 
dense-vector multiplication operation. For each $X_1[i], i \in \{1..n_1\}$, this phase gathers the minimum 
distances from $Z$ based on the positions of the words in $X_1[i]$ and then computes a dot-product with $F_{1,i}$.
Therefore, the overall functionality is equivalent to the quadratic-complexity RWMD given in Section~\ref{sec:complexity}.
Note that the second phase computes distances across all histograms in $X_1$ and a single histogram from $X_2$ in parallel, 
hence its time complexity is $O(nh)$. A main advantage of our method is that the relatively high cost 
of the first phase is amortized over a large number of histograms $X_1[i], i \in \{1..n_1\}$ in the second phase. 

The overall complexity of the linear complexity RWMD (LC-RWMD) algorithm when comparing a single histogram from 
$X_2$ against all histograms in $X_1$ is then $O(vhm + nh)$. The overall complexity when comparing all histograms
of $X_2$ against all histograms of $X_1$ is $O(nvhm+n^2h)$. When $n$ and $v$ are of the same complexity,
the overall complexity becomes $O(n^2hm)$. Even though this complexity still grows quadratically with the 
number of histograms (i.e., $n$), it grows linearly with the size of the histograms (i.e., $h$). 

The algorithm described so far computes the costs of moving histograms 
$X_1[i], i \in \{1..n_1\}$ to histograms $X_2[j], j \in \{1..n_2\}$. Assume that these results are stored
in an $n_2 \times n_1$ matrix $D_1$. To achieve a tight lower bound for WMD, the costs of moving 
the histograms $X_2[j], j \in \{1..n_2\}$ to $X_1[i], i \in \{1..n_1\}$, also have to be computed. Therefore, 
after computing $D_1$ in full, we swap $X_2$ and $X_1$, and run the LC-RWMD algorithm once more, which 
produces an $n_1 \times n_2$ matrix $D_2$. The final distance matrix $D$ is produced by computing the maximum 
values of the symmetric entries in $D_1$ and $D_2$ (i.e., $D=\max(D_1,D_2^T)$). 

When $O(n)=O(v)$, our approach reduces the average complexity of computing RWMD 
between a pair of histograms from $O(h^2m)$ to $O(hm)$. Such a reduction 
makes LC-RWMD orders of magnitude faster than the straightforward RWMD.
When $O(n) \neq O(v)$, the reduction in complexity is given by $min(nh/v,hm)$, 
which is typically determined by the first term, i.e., $nh/v$. Thus, LC-RWMD 
has significant advantages when the size of the histograms $h$ is large or the 
number of histograms $n$ is larger than the size of the vocabulary (i.e., $v$). 
Therefore, an important optimization we have in our LC-RWMD implementation is to eliminate 
the words that do not appear in $X_1$ from the vocabulary. Similarly, when $X_1$ 
and $X_2$ are swapped, the words that do not appear in $X_2$ are eliminated.

\medskip
\noindent
\textbf{Many-to-many LC-RWMD:} The LC-RWMD technique described so far compares a single histogram from one set with all 
histograms from another set. Although this approach is very fast when both sets are very large, 
it does not offer similar benefits when one or both sets are small in size. In particular, 
comparing an $X_2[j], j \in \{1..n_2\}$ with all $X_1[i], i \in \{1..n_1\}$ may not use 
the available compute resources in full if $n_1$ is not large enough. To circumvent this 
problem, we have developed a many-to-many implementation of LC-RWMD, wherein several 
histograms from $X_2[j], j \in \{1..n_2\}$ are compared with all $X_1$ concurrently. 

Assume, for simplicity, that all $n_2$ histograms from $X_2$ are compared with 
all $n_1$ histograms from $X_1$ in parallel. In this case, the first phase of the LC-RWMD 
algorithm computes $E \circ T_2$, where all $T_2[j], j \in \{1..n_2\}$ are combined in 
a single matrix $T_2$ with $O(n_2h)$ rows and $m$ columns. Afterwards, row-wise minima of 
$E \circ T_2$ are computed separately for each $j \in \{1..n_2\}$, which produces a $Z$ matrix 
of dimension $v \times n_2$. Next, the second phase of the LC-RWMD algorithm simply mulltiplies 
the $n_1 \times v$ sparse matrix $X_1$ with $Z$ to produce an $n_1 \times n_2$ distance matrix. 
In practice, it is not necessary to compute all $n_1 \times n_2$ distances in a single pass of the algorithm, but
the set $X_2$ can be divided into smaller batches that can be compared with all $X_1$ concurrently.

\section{GPU implementations} \label{sec:parallel}

We have developed GPU-accelerated and distributed implementations of our linear-complexity RWMD algorithm as well as the more straightforward quadratic-complexity approach, and demonstrated excellent scalability for both.

\begin{figure}[t!]
  \centering%
  \begin{tabular}{@{}c@{}}
  \includegraphics*[width=1.0\linewidth]{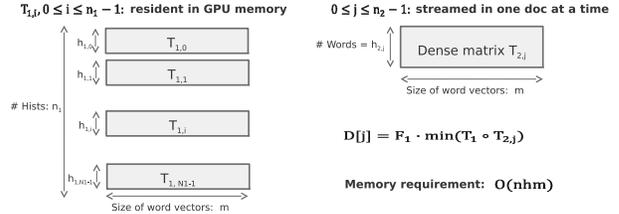}
  \end{tabular}
  \vspace{-0.6in}
  \caption{Mapping the Quadratic-Complexity RWMD method to GPU.}
  \vspace{-0.05in}
   \label{fig:rwmd-gpu}
\end{figure}

Figure~\ref{fig:rwmd-gpu} depicts the GPU implementation of the quadratic-complexity RWMD method detailed in Section~\ref{sec:complexity}. Here, we parallelize the computation of a complete column of the pairwise 
distance matrix $D$ by storing all $T_{1,i}, i \in \{1..n_1\}$ matrices in the GPU memory, and by performing pairwise RWMD computations across all $T_{1,i}, i \in \{1..n_1\}$ and a single $T_{2,j}, j \in \{1..n_2\}$ 
at once. The RWMD computations are performed using an adaptation of the implementation given in Section~\ref{sec:complexity}. The Euclidean distance computations across all $T_{1,i}, i \in \{1..n_1\}$ matrices and a 
single $T_{2,j}, j \in \{1..n_2\}$ are performed by combining all $T_{1,i}, i \in \{1..n_1\}$ in a single matrix $T_1$ with $O(nh)$ rows and $m$ columns, and by using NVIDIA's CUBLAS library for multiplying it with
$T_{2,j}$. The row-wise and column-wise minimum operations are performed using NVIDIA's Thrust library. The dot-product operations are performed using CUBLAS again. Such an implementation requires $O(nhm)$ space in 
the GPU memory because each $T_{1,i}$ matrix requires $O(hm)$ space, and we store $O(n)$ such matrices simultaneously in the GPU memory.

\begin{figure}[t!]
  \centering%
  \begin{tabular}{@{}c@{}}
  \includegraphics*[width=1.0\linewidth]{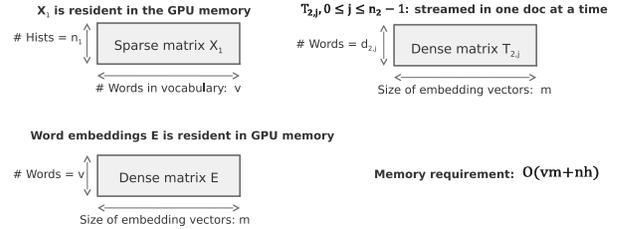}
  \end{tabular}
  \vspace{-0.55in}
  \caption{Mapping the Linear-Complexity RWMD method to GPU.}
  \vspace{-0.1in}
   \label{fig:lcrwmd-gpu}
\end{figure}

Fig.~\ref{fig:lcrwmd-gpu} shows the GPU implementation of the linear complexity RWMD method detailed in Section~\ref{sec:complexity}. In the first phase, our implementation uses the CUBLAS library for computing 
Euclidean distances between $E$ and a single $T_{2,j}, j \in \{1..n_2\}$, and then the Thrust library for computing row-wise minima. In the second step, we use NVIDIA's CUSPARSE library for multiplying
$X_1$ by the result of the first phase. Unlike the quadratic-complexity implementation, the linear-complexity RWMD does not store $T_{1,i}, i \in \{1..n_1\}$, matrices in the GPU memory. Instead of allocating
the dense $T_1$ matrix that stores embedding vectors for all words in all histograms in $X_1$, LC-RWMD simply stores the sparse matrix $X_1$, which requires $O(nh)$ space, 
and the embedding vectors for the complete vocabulary (i.e., $E$), which requires $O(vm)$ space. Therefore, the overall space complexity of LC-RWMD is $O(nh+vm)$. When $O(v)=O(n)$, 
the overall space complexity becomes $O(nh+nm)$. As a result, the space complexity of LC-RWMD is smaller than that of the quadratic-complexity RWMD by a factor of $min(h,m)$.
In conclusion, LC-RWMD reduces not only the time complexity by a factor of $h$ with respect to the straightforward RWMD, but also the space complexity by a similar factor.
Table~\ref{tab:complexity} summarizes these results.

\begin{table}[t!]
\caption{Complexity Comparison}
\vspace{-0.15in}
\label{tab:complexity}
\begin{center}
\begin{tabular}{ccc}
  \hline
              & Time Complexity & Space Complexity \\ 
  \hline
  LC-RWMD     & $O(nvhm+n^2h)$     & $O(nh+vm)$ \\
  RWMD        & $O(n^2h^2m)$       & $O(nhm)$ \\ 
  Reduction   & $O(min(nh/v,hm))$  & $O(min(nh/v,m))$ \\ 
  \hline
  \vspace{-0.35in}
\end{tabular}
\end{center}
\end{table}

All our algorithms are data-parallel and can easily be distributed across several GPUs. 
Spreading either $X_1$ or $X_2$ across several GPUs is sufficient. The only function that may require communication
in distributed setting is the top-$k$ computation. However, the associated communication cost is typically 
marginal compared with the cost of computation.

\section{Performance Evaluation} \label{sec:performance}

Table~\ref{tab:datasets} summarizes the characteristics of the
datasets used in our experiments in terms of the number of 
documents ($n$) and the average number of unique words per 
document excluding the stop-words, which indicates the average 
histogram size ($h$). Both datasets are proprietary 
and used in our production systems for news classification. 

\begin{table}[h!]
\caption{Characteristics of the datasets used in experiments}
  \vspace{-0.1in}
\label{tab:datasets}
\begin{center}
\begin{tabular}{lccc}
  \hline
  Dataset      	  & $n$    & $h$ (average) & $v_{e}$\\
  \hline
  Set 1           & 1M  & 107.5 & 452,058 \\
  Set 2           & 2.8M & 27.5  & 292,492 \\ 
  \hline
  \vspace{-0.2in}
\end{tabular}
\end{center}
\end{table}

Our algorithms require two datasets for comparison. 
The first set $X_1$ is resident (i.e., fixed) and the 
second set $X_2$ is transient (i.e., unknown). In our 
experiments, we treat the datasets given in Table~\ref{tab:datasets} 
as resident sets. Given a second set of documents 
(i.e., $X_2$), we compare it with $X_1$. In typical 
use cases, $X_1$ and $X_2$ are completely independent sets, 
e.g., one is the training set and the other is the test set.
However, in the experiments presented, we define $X_2$ 
to be a randomly sampled subset of $X_1$ for the sake of simplicity.

The experiments presented in this section
use embedding vectors generated by \textit{word2vec}~\cite{MikolovCorr2013} on Google News, 
where the vocabulary size is three million words (i.e., $v=3,000,000$),
and each word vector is composed of 300 single-precision 
floating-point numbers (i.e., $m=300$). 
However, the number of embedding vectors the LC-RWMD algorithm 
stores in the GPU memory is given by the number of unique words 
that exist in the resident dataset. We call this number 
$v_{e}$, and show the respective values in Table~\ref{tab:datasets}. 

We deployed our algorithms on a cluster of four IBM POWER8+ nodes, 
where each node had two-socket CPUs with 20 cores and 512 GB of CPU memory. 
\footnote{IBM is a trademark of International
Business Machines Corporation, registered in many jurisdictions worldwide. 
Other product or service names may be trademarks or service marks of IBM or other companies.}
Each node had four NVIDIA Tesla P100 GPUs attached 
via NVLINK interfaces. A P100 GPU has 16 GB of memory, half of which we 
used to store the resident data structures, and the rest for the temporary
data structures. The POWER8+ nodes were connected via 100 Gb Infiniband 
interfaces. All our code is written in C++. We used version 14.0.1 of 
IBM's XL C/C++ compiler to build our code, CUDA 8.0 to program 
the GPUs, and MPI for inter-node and inter-process communication. 

\begin{figure}[t!]
  \vspace{-0.2in}
  \centering%
  \begin{tabular}{@{}c@{}}
  \includegraphics*[width=0.8\linewidth]{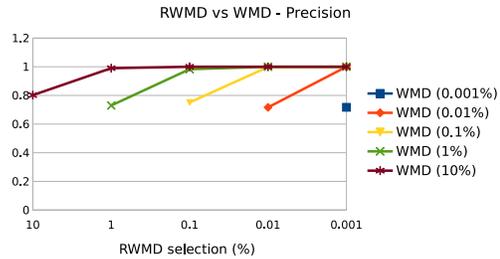}
  \end{tabular}
  \vspace{-0.2in}
  \caption{Amount of overlap between top-$k$ results of RWMD and WMD.}
   \label{fig:precisionrwmd}
\end{figure}

\begin{figure}[t!]
  \vspace{-0.2in}
  \centering%
  \begin{tabular}{@{}c@{}}
  \includegraphics*[width=0.8\linewidth]{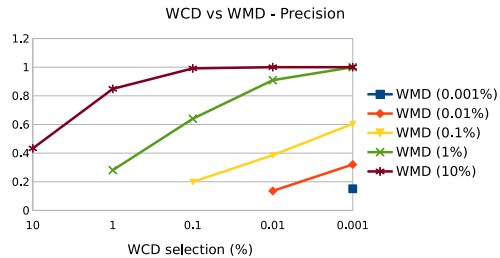}
  \end{tabular}
  \vspace{-0.2in}
  \caption{Amount of overlap between top-$k$ results of WCD and WMD.}
  \vspace{-0.15in}
   \label{fig:precisionwcd}
\end{figure}

Figure~\ref{fig:precisionrwmd} shows the ratio of overlap 
between the top-$k$ results of WMD and the top-$k$ results of RWMD. 
The line labeled \emph{WMD (1\%)} indicates the ratio of the 
top-$k$ results of the RWMD that overlap with the top 1\% 
results of WMD, where $k$ is determined based on the percentage 
given on the $x$-axis. Figure~\ref{fig:precisionrwmd} shows that the 
ratio of overlap varies between 0.72 and 1 for RWMD, and proves 
that RWMD is a high-quality aproximation of WMD. 
The same analysis is performed between WCD and WMD in Fig.~\ref{fig:precisionwcd},
where we observe overlap ratios that are as low as 0.13, 
indicating that WCD is not a high-quality approximation of WMD.

\begin{figure}[t!]
  \vspace{-0.15in}
  \centering%
  \begin{tabular}{@{}c@{}}
  \includegraphics*[width=0.8\linewidth]{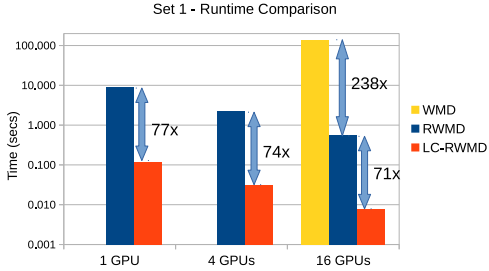}
  \end{tabular}
  \vspace{-0.15in}
  \caption{Time to compare a single document with 1M documents.}
   \label{fig:blekkospeedup}
\end{figure}

\begin{figure}[t!]
  \vspace{-0.15in}
  \centering%
  \begin{tabular}{@{}c@{}}
  \includegraphics*[width=0.8\linewidth]{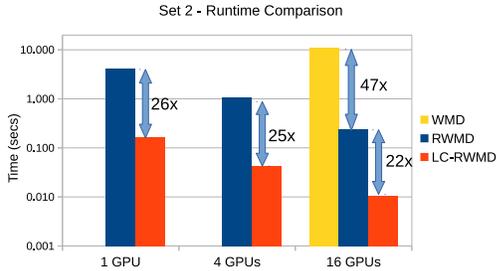}
  \end{tabular}
  \vspace{-0.10in}
  \caption{Time to compare a single document with 2.8M documents.}
  \vspace{-0.15in}
   \label{fig:ciqspeedup}
\end{figure}

Figure~\ref{fig:blekkospeedup} and Figure~\ref{fig:ciqspeedup}
compare LC-RWMD, the quadratic-complexity RWMD, and WMD in terms of the runtime 
performance. Figure~\ref{fig:blekkospeedup} shows the average time to compute 
the distance between a single transient histogram and one million resident 
histograms from Set 1. Similarly, Fig.~\ref{fig:ciqspeedup} 
shows the average time to compute  the distance between a single transient 
histogram and 2.8 million resident histograms from Set 2.
We observe that LC-RWMD is faster than straightforward RWMD by approximately a 
factor of $h$. Notably, when running LC-RWMD on 
a single GPU, the time to compute one million distances is around 120 ms 
for Set 1. For Set 2, which has smaller histograms, 
2.8 million distances can be computed in approximately only 160 ms. 

Both the straightforward RWMD and the LC-RWMD benefit from parallelization 
on multiple GPUs. Both are data-parallel algorithms and demonstrate linear 
scaling as we increase the number of GPUs used. Notably, the time to compare
one transient document against all resident documents is approximately 
8 ms for Set 1 and 10 ms for Set 2  when running LC-RWMD on 16 GPUs. 
We distribute either the resident dataset or the transient dataset across multiple GPUs. 
In the case of the straightforward RWMD, the storage requirements are much 
higher per histogram, and hence, fewer histograms can be fit into 
the GPU memory. When the complete resident set does not fit into the GPU 
memory, the resident histograms are copied into the GPU memory in several 
batches, which creates additional overhead. To minimize this overhead, 
we distributed the resident set when multiple GPUs were available. For instance, 
when using 16 GPUs, the complete Set 2 data fits into the distributed GPU 
memory, and the copy overheads are completely eliminated, which results in 
a super-linear speedup for the straightforward RWMD as shown in Fig.~\ref{fig:ciqspeedup}. 
The LC-RWMD uses more compact data structures, enabling the complete resident 
set to be stored in the memory of a single GPU. When several GPUs are available, 
it is advisable to replicate the smaller set and distribute the larger set for LC-RWMD.

Figure~\ref{fig:blekkospeedup} and Figure~\ref{fig:ciqspeedup} also
show the runtime results for WMD, which are approximately two orders 
of magnitude higher than those of the quadratic-complexity RWMD. 
In this work, we developed a parallel WMD implementation that
distributes the resident dataset across several CPU processes, 
wherein each CPU process owns a dedicated GPU that performs the 
Euclidean distance and RWMD computations. In addition, we
implemented the pruning algorithm described in Section~\ref{sec:complexity} 
to significantly reduce the number of EMD computations. 
Each CPU process then runs the state-of-the-art EMD library~\cite{emd} 
on its share of the resident data independently.
Note that in the pruning algorithm, the top-$k$ results computed by 
RWMD helps derive a cut-off value. Figure~\ref{fig:blekkospeedup} 
and Figure~\ref{fig:ciqspeedup} show the average runtime results 
for WMD for $k=128$ when using 16 CPU processes and 16 GPUs 
controlled by those processes. Of course, using a smaller $k$
value reduces the time to compute WMD. For instance, when
using $k=16$, the time our algorithms took to compute the WMD 
dropped by a factor of three for each set. Nevertheless,
when computing WMD, we had to use very small transient sets 
(a few hundred documents from Set 1 and a few thousand documents 
from Set 2, randomly sampled) to make the WMD computation tractable. 

\begin{figure}[t!]
  \begin{tabular}{@{}c@{}}
  \includegraphics*[width=0.92\linewidth]{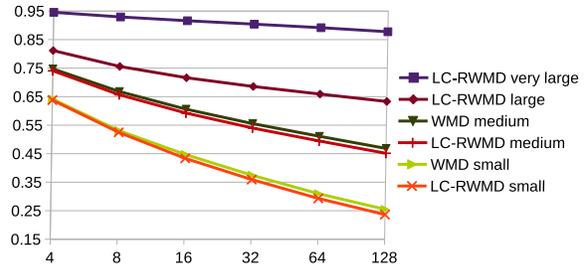}
  \end{tabular}
  \vspace{-0.05in}
  \caption{WMD vs LC-RWMD: Precision at top-$k$ for Set 2.}
  \vspace{-0.12in}
   \label{fig:ciqprecision}
\end{figure}

Figure~\ref{fig:ciqprecision} evaluates the suitability of WMD and 
LC-RWMD for $k$-nearest-neighbors classification, where the $x$-axis 
indicates $k$ and the $y$-axis indicates the precision at top-$k$. 
Here we used the fully labeled Set 2, where we organized the labels 
into four subsets: those that have 300 to 1000 examples are part of 
the \textit{small} subset, those that have 1000 to 10,000 examples 
are in the \textit{medium} subset, those that have 10,000 to 100,000 
examples are in the \textit{large} subset and those that have 100,000 
to one million examples are in the \textit{very large} subset. For 
each document, we computed the fraction of the documents that have 
the same label in its top-$k$ list. For each $k$, we averaged 
these results within the same label, and then computed the 
geometric mean across different labels within the same subset.
The results indicate that the precision achieved by LC-RWMD is
very close to that of WMD for small sets. However, for large 
sets, WMD becomes intractable. The results also indicate that 
the precision is a function of the size of the datasets. Thus,
there is a real need for scalable algorithms, such as LC-RWMD.

\section{Related Work} \label{sec:relatedwork}

Several methods have been proposed to achieve lower bounds for EMD that will reduce the time to answer nearest-neighbor queries~\cite{assent2008efficient,  xu2010efficient, ruttenberg2011indexing, wichterich2008efficient, xu2016emd, huang2016heads, huang2014melody}. 
Those lower bounds can be used in multi-step query architectures in which an initial filter step is followed by a refinement step. 
Significant scalability has been observed, especially when the lower bounds are such that they can be used by an index structure. 
In these approaches, the speedup is achieved by an efficient pruning of the potential candidates. 
Nevertheless, in very large sets, the number of remaining candidates that the EMD has to be computed on is still big enough to be prohibitive.
Alternatively, a compressed representation of the documents can be used to compute EMD fast, but approximately~\cite{Uysal2016approximation}. 
Another approximate EMD computation algorithm has been proposed in the image-processing domain, which notably achieves linear time complexity~\cite{ShirdhonkarJ08} using a wavelet-based approach. 
However, whether the algorithm proposed in~\cite{ShirdhonkarJ08} can be extended to compute WMD is an open research problem.
 
To the best of our knowledge, the above-mentioned techniques have not been directly used in the computation of the WMD. However, some lower bounds of the WMD have been introduced in~\cite{KusnerSKW15}. 
For instance, the Word Centroid Distance provides a fast and scalable lower bound and can be used to facilitate the pruning in a nearest-neighbor query. 
RWMD is a tighter lower bound of WMD, but its quadratic time complexity presents scalability issues (see Section~\ref{sec:complexity}).	

Recently, Huang et al. proposed a supervised version of the WMD algorithm that can improve the accuracy of distance computation with respect to the original unsupervised WMD algorithm~\cite{HuangGKSSW16}. 
The LC-RWMD algorithm proposed by this work is also unsupervised, but can be extended to support supervision. The relaxation technique used by Huang et al. in~\cite{HuangGKSSW16} relies on Cuturi's Sinkhorn Distance algorithm~\cite{Cuturi13}, which has quadratic time complexity. Therefore, it is not as scalable as the LC-RWMD algorithm.

\section{Conclusion} \label{sec:conclusion}

The Relaxed Word Mover's Distance (RWMD) was proposed by Kusner et al. in their 
seminal paper as a tight lower bound for the popular Word Mover's Distance. 
However, Kusner et al. proposed a quadratic-complexity implementation of RWMD. 
In this work, we show that the Relaxed Word Mover's Distance can be implemented 
in linear time on average when computing distances across large sets of documents.
We then show a practical implementation of our method, which maps well onto commonly
used dense and sparse linear algebra routines, and can be executed efficiently
on GPUs. In addition, we demonstrate that our implementations can be efficiently scaled 
out across several GPUs and exhibit a perfect strong or weak scaling behavior.

\section{Acknowledgement}
We thank Dr. Nikolas Ioannou from IBM Research for his technical comments and 
Ms. Charlotte Bolliger from IBM Research for her language-related corrections.

\bibliographystyle{unsrt}

\end{document}